\pdfoutput=1
\documentclass[%
reprint,
superscriptaddress,
showkeys,
 amsmath,amssymb,
 aps,
]{revtex4-2}
\pdfoutput=1
\usepackage{graphicx}
\usepackage{dcolumn}
\usepackage{bm}
\usepackage{hyperref}
\usepackage[section]{placeins}
\hypersetup{
     colorlinks   = true,
     linkcolor    = blue,
     citecolor    = blue,
     urlcolor     = blue
}

\newcommand{\orcid}[1]{\href{https://orcid.org/#1}{\includegraphics[width=10pt]{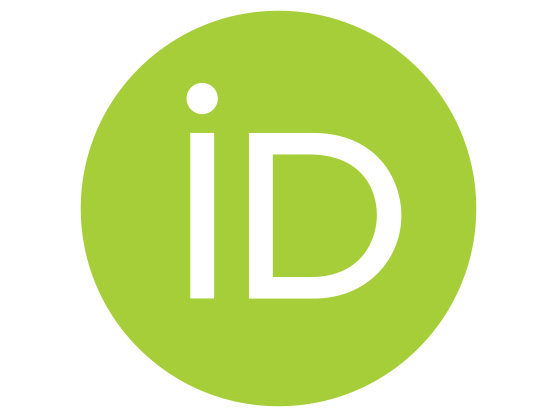}}}

\begin{document}

\pdfoutput=1

\preprint{APS/123-QED}

\title{Detecting and Denoising Gravitational Wave Signals from Binary Black Holes using Deep Learning}
%
\author{Chinthak Murali \orcid{0000-0001-8475-7833}}
\affiliation{Department of Physics, The University of Texas at Dallas, Richardson, TX 75080, USA}

\author{David Lumley \orcid{0000-0003-4342-4551}}
\affiliation{Department of Physics, The University of Texas at Dallas, Richardson, TX 75080, USA}
\affiliation{Department of Geosciences, The University of Texas at Dallas, Richardson, TX 75080, USA}

\begin{abstract}

 We present a convolutional neural network, designed in the auto-encoder configuration that can detect and denoise astrophysical gravitational waves from merging black hole binaries, orders of magnitude faster than the conventional matched-filtering based detection that is currently employed at advanced LIGO (aLIGO). The Neural-Net architecture is such that it learns from the sparse representation of data in the time-frequency domain and constructs a non-linear mapping function that maps this representation into two separate masks for signal and noise, facilitating the separation of the two, from raw data. This approach is the first of its kind to apply machine learning based gravitational wave detection/denoising in the 2D representation of gravitational wave data. We applied our formalism to the  first gravitational wave event detected, GW150914, successfully recovering the signal at all three phases of coalescence at both detectors. This method is further tested on the gravitational wave data from the second observing run ($O2$) of aLIGO, reproducing all binary black hole mergers detected in $O2$ at both the aLIGO detectors. The Neural-Net seems to have uncovered a pattern of 'ringing' after the ringdown phase of the coalescence, which is not a feature that is present in the conventional binary merger templates. This method can also interpolate and extrapolate between modeled templates and explore gravitational waves that are unmodeled and hence not present in the template bank of signals used in the matched-filtering detection pipelines. Faster and efficient detection schemes, such as this method, will be instrumental as ground based detectors reach their design sensitivity, likely to result in several hundreds of potential detections in a few months of observing runs. 
\end{abstract}

\maketitle


\section{\label{sec:level1}INTRODUCTION}
The first direct detection of a gravitational wave (GW), GW150914 in 2015 has opened a new window into observing the universe \cite{150914, properties}. In the first three observing runs ($O1$, $O2$ and $O3$) of aLIGO \cite{ligo} and the European VIRGO detector \cite{virgo} over the span of 27 months, a total of 90 CBC (Compact Binary Coalescence) events with an astrophysical origin probability $p_\text{astro}>0.5$ have been detected, as reported in the Gravitational Wave Transient Catalogues, GWTC-1 \cite{gwtc1}, GWTC-2 \cite{gwtc2} and GWTC-3 \cite{gwtc3}. This includes 85 binary black hole (BBH) mergers, four neutron star black hole (NSBH) mergers  and one binary neutron star (BNS) merger \cite{neutron}. Hundreds of more detections are expected in the scheduled $O4$ run of aLIGO with an enhanced sensitivity \cite{sensitivity} and being joined by the Japanese underground detector KAGRA \cite{kagra1,kagra2}. With a number of ground based GW detectors under construction \cite{india,et}, the demand for more efficient and reliable data processing techniques are only getting higher.

aLIGO uses matched-filtering based search to identify the presence of a GW in the data stream \cite{matched}. In this process, an enormous template bank of signals are cross-correlated with several months of detector data output, creating an alert when the matched-filter signal to noise ratio (SNR) exceeds the detection threshold of the detection pipeline. Once the presence of a GW is identified in the time segment, traditional signal processing techniques are applied to enhance the signal and suppress the noise, to retrieve the signal from the data stream. The template with the highest SNR is then used to characterize the merger event. A merging BBH system is characterized by eight intrinsic parameters, the two masses and the two spins (magnitudes and directions) and the extrinsic parameters: the luminosity distance, right ascension, declination, polarization, inclination, coalescence time, coalescence phase and two parameters of eccentricity. Sampling a fifteen dimensional parameter space to characterize each individual GW waveform and performing matched-filter analysis with each data segment is not a trivial computational task \cite{paramspace}.  

The recent advances in machine learning (ML), namely Deep Learning (DL), can help us navigate this intense computational problem. DL is an extremely powerful machine learning technique that can learn very complex features and functions through neural networks~\cite{supervised, dll}. Convolutional Neural Networks (CNN) is a special class of neural networks that performs convolution operations by means of kernel filters to the input data. During the training process, the CNN assigns weights to the filters that optimally extracts various features from the input data. As opposed to matched-filtering based search, in a DL based search, all the intensive computations are performed only during the training stage of the network, which is a one time process \cite{george1}. This procedure can also interpolate and in some cases extrapolate between waveform templates, making it more robust in detecting GW signals without necessarily training the network on a 15D parameter space. DL based search also opens up the possibility of detecting GW signals that are outside the realm of theoretical template banks currently modeled using Numerical Relativity (NR). Because of the ability of the neural networks to process data and detect the GWs orders of magnitude faster, it can be effectively combined with the matched-filtering based detections to enhance the confidence of a given detection and also to efficiently target data segments with positive detections for a subsequent matched-filter based analysis.

Several ML based analysis of GW data have appeared in the literature over the past few years, beginning with glitch classification and subtraction \cite{glitch1, glitch2, glitch3, glitch4, glitch5, glitch6}. DL methods are first applied to direct detection of GWs by George and Huerta using simulated aLIGO noise \cite{george1} and further extended to real aLIGO noise, resulted in detecting the presence of the first GW, GW150914 \cite{george2} in the data stream. This work has inspired several attempts to identify and locate GWs in real aLIGO data \cite{ml1, ml2, ml3, ml4, ml5, ml6} and many other ML based studies focused on parameter estimation of real GWs \cite{MLL1, MLL2, MLL3, MLL4} followed. Denoising GWs using DL was applied in \cite{dl1}, and the proposed denoising scheme was able to extract four GW events (three from Hanford, one from Livingston) with high signal overlaps. In \cite{dl2, dl3} the authors proposed a denoising auto-encoder, based on Recurrent Neural Networks (RNN) to denoise GWs. Most recently, \cite{ml-gw} used 1D CNN in an auto-encoder configuration to denoise many of the GWs (three events from $O1$ and three events from $O2$), using whitened data from both the detectors. 

 Although George and Huerta \cite{george1} in the foundational article suggested that using 2D data in the context of GW detection is sub-optimal because of the extremely weak signal strength characteristic of GWs, we are using 2D representation of the data in order to separate signal from the noise. In our denoising scheme, we use 2D CNN architecture on raw detector data, with un-whitened signal templates and noise. This approach, to the best of our knowledge is the first attempt to detect and denoise GWs from raw data in 2D representation.

In this paper we present a DL framework to detect and denoise GWs from raw strain data from the aLIGO detectors. The detector noise at both aLIGO detectors are highly non stationary and non-Gaussian, with very high noise dominating both ends of the frequency spectrum \cite{noises}. While ambient seismic noise (from ocean, traffic, earthquakes, etc.) dominate the low frequency end of the noise \cite{seismic}, photon shot noise dominate the high frequency regime. The real GW signals are extremely weak and are deeply buried inside the detector noises and occupy the same frequency band as the detector noise, making it impossible to separate from the noise using traditional filtering methods. 

In this method, we implement a time-frequency denoising technique used in seismic denoising, which successfully separated earthquake signals from ambient noise in 2019 \cite{denoise} using DL. The noisy strain data is first transformed into time-frequency domain using Short Time Fourier Transform (STFT), rendering the one dimensional time series data into two dimensions, represented by Fourier coefficients. In the Fourier domain, the coefficients associated with noise are attenuated to enhance the signal coefficients. These modified Fourier coefficients are then transformed back in to the time domain using inverse Fourier transform and thereby reconstructing the GW chirp signal in the time domain. The underlying idea is to promote a sparse representation of the signal in the time-frequency domain where the signal can be represented by a sparse set of features which makes the separation of the signal from the noise easier in the Fourier domain. This method is especially useful when the GW signal and the detector noise occupy the same frequency band, making the filtering techniques virtually impractical.

\section{Method}

The key here is to find a mapping function that can appropriately find a threshold to suppress the coefficients corresponding to the noise in the time-frequency domain, hence enhance signal separation. These functions are highly non-linear and are hence difficult to construct mathematically for the the GW problem. That is where DL techniques can be incredibly effective, which learns to build a high dimensional, non-linear mapping function from the data alone during the training process. This Neural-Net learns the sparse representation of the data in the time-frequency domain, and builds a high-dimensional, non-linear mapping function which maps these representations into two masks, one for the GW signal and another one for all the noises.

The raw GW data from the detector $d(t)$ undergoes STFT and is represented in the time-frequency domain as $D(t, f)$, which is a combination of the GW signal $S(t, f)$ and all the noises $N(t, f)$, 
\begin{equation}
    D(t, f) = S(t, f) + N(t, f).
\end{equation}
The idea is to construct the mapping functions that can successfully map the detector data into a representation of the signal and a representation of the noise separately. This mapping can be accomplished through a soft thresholding in the sparse representation where the threshold is estimated by assuming a Gaussian distribution of noise \cite{donoho}.

From here we construct two individual masks $M_S(t, f)$ and $M_N(t, s)$ which act as the mapping functions for the signal and noise respectively, and are given by, 

\begin{equation}
    M_S(t, f) = \frac{1}{1+\frac{|N(t, f)|}{|S(t, f)|}},
\end{equation}    
    
\begin{equation}
   M_N(t, f)= \frac{\frac{|N(t, f)|}{|S(t, f)|}}{1+\frac{|N(t, f)|}{|S(t, f)|}}.
\end{equation}

These masks are the targets for the supervised learning problem, and they are constructed during the training process from the training data. The Neural-Net is trained to construct these masks at the output layer, for every single data segment that is fed into the network as part of the training data. Once these masks are constructed, they can be multiplied with the 2D detector data $D(t, f)$ to reconstruct the signal and noise as follows.

\begin{equation}
    M_S(t, f) \times D(t, f) \sim S(t, f),
\end{equation}

\begin{equation}
    M_N(t, f) \times D(t, f) \sim N(t, f).
\end{equation}

These reconstructed signal and noise are inverse transformed back into the time domain to reproduce the one dimensional time series of signal and noise,
\begin{equation}
    STFT^{-1} (S(t, f)) = S(t),
\end{equation}
\begin{equation}
    STFT^{-1} (N(t, f)) = N(t). 
\end{equation}

The denoising process aims to find the true GW signal $\hat{S}$ from the data by minimising the mean squared error between the true signal and the estimated signal, calculated as,

\begin{equation}
    \text{Error}= ||\hat{S}-S||^2.
\end{equation}

\begin{figure*}[t]
\includegraphics[width=\textwidth]{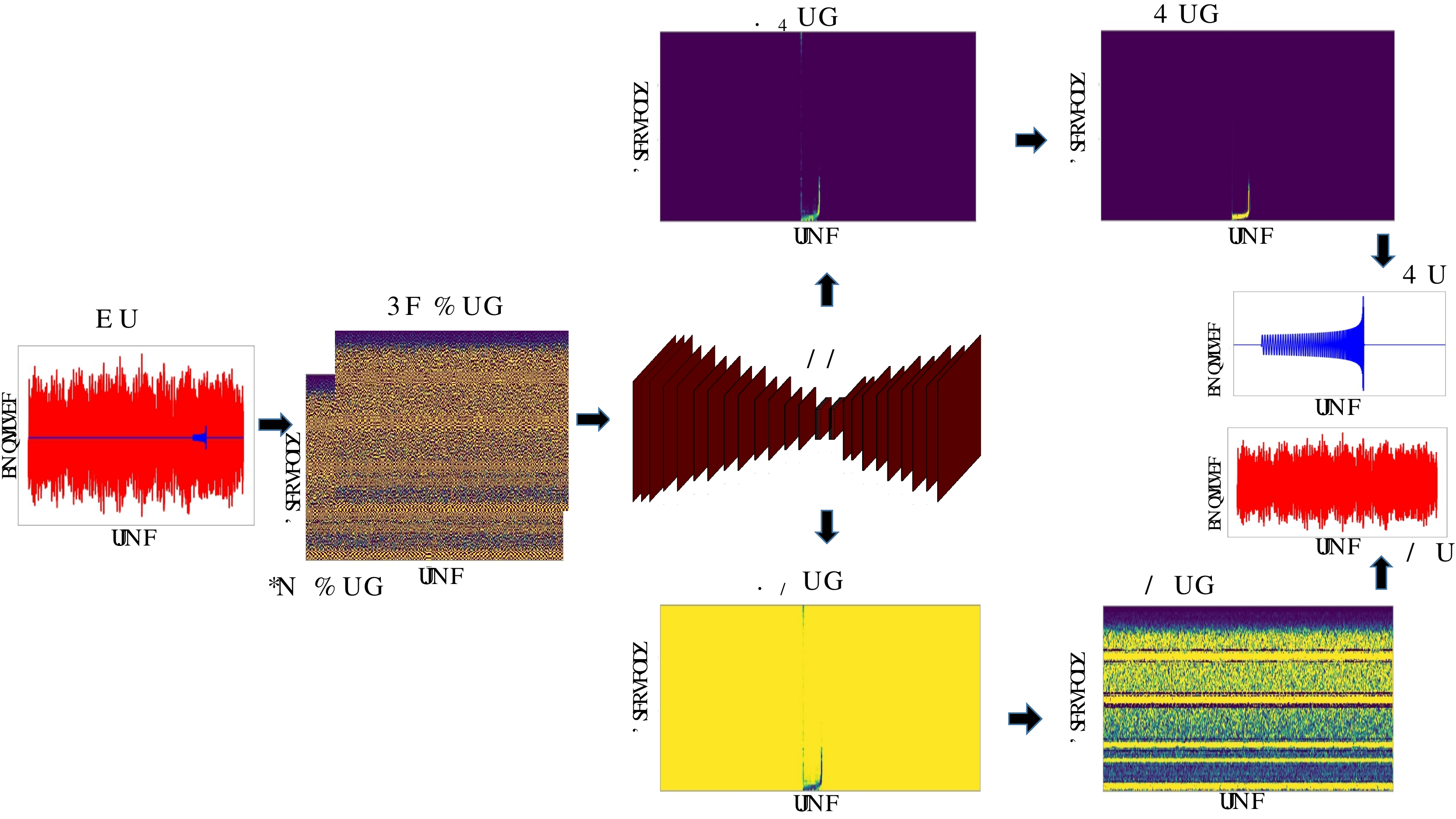}
\caption{Denoising flow diagram. Input is the data in the time domain, which undergoes STFT to produce the real and imaginary parts of the data, which are inputs to the Neural-Net. The Neural-Net generates masks for the signal and the noise, which are multiplied with the data itself to create the 2D representation of signal and noise separately. An inverse STFT reconstructs the GW signal and noise in the time domain.}
\label{flow}
\end{figure*}

The masks $M_S(t, f)$ and $M_N(t, f)$ have the same dimension as the input data $D(t, f)$ and take values between $0$ and $1$ and are mutually exclusive $(M_N = 1-M_S)$. 
The entire process of signal and noise separation is presented as a sequence in Figure \ref{flow}.

\begin{figure*}[t]
\includegraphics[width=\textwidth, height=10cm]{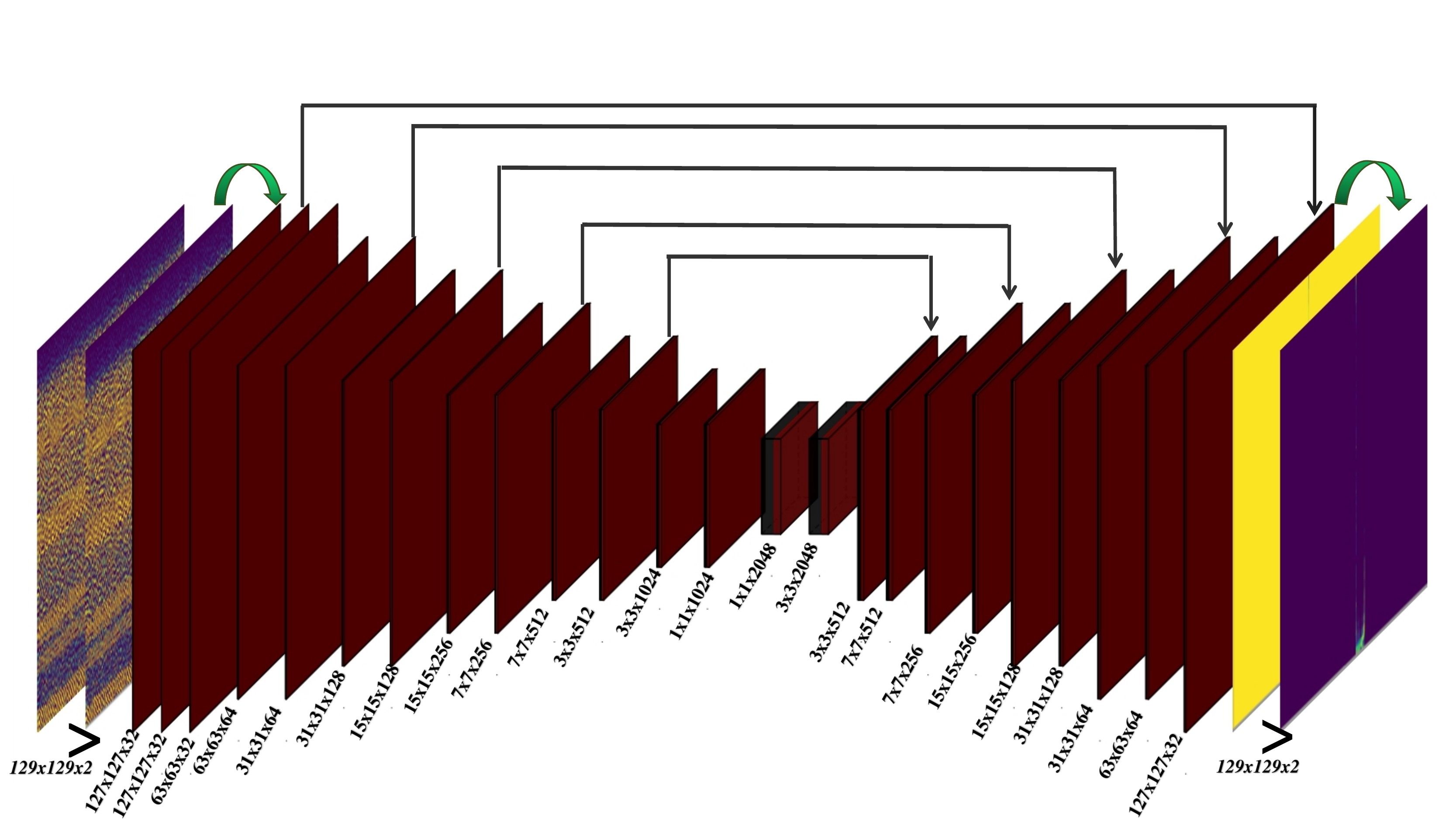}
\caption{Neural-Net architecture. Inputs are the real and imaginary components of the Fourier coefficients, and outputs are the masks $M_S$ and $M_N$. Each red rectangle is a 2D convolutional layer, with dimensions labelled at the bottom as ``frequency bins $\times$ time points $\times$ channels". Overhead arrows are skip connections between the encoder and the decoder parts of the network.}
\label{nn}
\end{figure*}

\subsection{Neural-Net Architecture}
Auto-encoders are a class of artificial neural networks that can learn to code patterns from unlabelled data (unsupervised learning), typically for dimensionality reduction \cite{autoencoder}. The encoder part of the auto-encoder maps the representation to a code and the decoder part reconstructs the coded representation. Because of the ability of auto-encoders to learn a sparse representation of the data, our Neural-Net is designed as a series of fully connected 2 dimensional convolutional layers with descending and then ascending sizes, like an auto-encoder, as shown in Figure \ref{nn}. Skip connections are used to improve convergence during the training process \cite{skip}, which are represented as over-head arrows connecting the encoder and decoder layers of same dimensions. In addition to improving the convergence, skip connections helped to minimize the signal leakage into the noise.

The input to the Neural-Net is through two channels : one takes the real part of the Fourier coefficients and the other takes the imaginary part of the Fourier coefficients. This enables the Neural-Net to learn from both the amplitude and phase information of the data. These inputs go through a series of 2D convolutional layers of constantly decreasing dimensions, where each layer uses a Rectified Linear Unit (ReLU) activation function and are Batch Normalized. The dimensions of the convolutional layers are reduced using a stride of $2 \times 2$ and the kernel size of each layer remains at $3 \times 3$. Each of the convolutional layers extract features from the data and learn to represent the data more and more sparsely as they go along the layers. At the bottleneck layer, we have the smallest and sparsest representation of the data possible and then it goes through the deconvolutional layers which generates a high-dimensional, non-linear mapping of this sparse representation into output masks. The output labels are the masks created for the signal ($M_S$) and for the noise ($M_N$). During the training process, the network learns to generate the masks that optimally separates signal from noise by minimizing a loss function, which is a binary cross-entropy loss function. A softmax normalized exponential function is used at the final layer to produce the output masks. Figure \ref{nn} shows the Neural-Net architecture which takes the inputs through two channels and generates two outputs, one of which is the familiar GW chirp signal in the time-frequency domain.

\subsection{Training Strategy}

The data used for training is the publicly available GW strain data from the Gravitational Wave Open Science Center (GWOSC \cite{gwosc}). These are continuous recordings at both the aLIGO detectors over the course of the first three observing runs. We used the data from $O1$ to train the network to detect the event GW150914 and data from $O2$ to train the network for the rest of the events. From a few hours of continuous noise recordings, we generated different realizations of the noise to create a larger noise set, to make the Neural-Net more adaptive to variations in noise. Data from GWOSC is sampled at $4096$Hz and are $4096$s long data segments. We down-sampled the data into $2048$Hz and divided them into eight second long data segments in the training phase. These eight second long data segments are pure noise where there is no known GW event reported so far, or any hardware injection, often used for calibration purposes. We applied a lower frequency cut-off of $30$Hz for this analysis as all the events detected so far are above this frequency. This noise is then combined with simulated GW signals, modeled using the optimised Effective One Body Numerical Relativity waveform SEOBNRv4\textunderscore opt \cite{opt} (an optimized version of SEOBNRv4 \cite{seobnr}) sampled at $2048$Hz with masses ranging from $5M\textsubscript{\(\odot\)}$ to $80M\textsubscript{\(\odot\)}$. For the analysis we assumed optimal orientation of both detectors with the event, also spins and eccentricities are assumed to be zero. This makes the analysis essentially two dimensional, where the signals are characterized by individual masses of the binary, which to the first order, captures the essence of the GW signal. 

The final one second of each merger is used as the GW signal that the Neural-Net is trained to identify and extract from the detector noise. For the entire subset of the signals used for the training purpose, there is enough signal in the inspiral, merger and ringdown phases, for effective patter recognition by the Neural-Net. The one second long signal is added to the eight second long noise by randomly time translating the signal within the noise to make the Neural-Net more robust to look for signals at different parts of the data, rather than being constrained at a single position where it is most likely to spot the signal. Another equivalent set is created with only noise and no GW signal, whose signal output is null and the noise output is the entire data. This helped reduce the False alarms considerably of the test set. A total of $22500$ waveforms are generated within the mass limits and combined with noise to make the first $22500$ samples in the training data. Another $22500$ samples are pure noise from the detectors with no signal injection. These $45000$ samples are normalized, shuffled and fed into the network for training and validation, over $10$ epochs and a small subset of that is used for testing. Both training and validation accuracy were $\sim 98 \%$.

Our training of the Neural-Net is a two step process. In step one, the simulated GW signals are combined with simulated noise of aLIGO detectors. We use the ``zerodetunedaLIGO" \cite{zero} model to simulate the noise. The Neural-Net is trained exclusively on modeled signal combined with modeled noise, and the parameters of the network are optimized solely based on this. The architecture and hyper-parameters that resulted in the maximum convergence and signal recovery is adopted as the optimal configuration of the Neural-Net. In step two, the weights of the Neural-Net from step one are transferred to the new Neural-Net whose inputs are created by combining simulated GW signals with real detector noise from both aLIGO detectors separately. This technique, called \texttt{transfer learning} has been shown to be very robust in enhancing convergence and improving the accuracy of neural networks in many areas, including GW detection \cite{george2}. Transfer learning has improved the performance of the Neural-Net by both reducing the false alarms and by improving signal extraction with higher overlap with the template of a given GW. Training the network with both simulated noise and real noise together did not improve the detection/denoising capability as transfer learning, even though in both cases the Neural-Net is essentially presented with the same training data set. 
As noted in \cite{training}, networks trained on lower SNR signals generally outperformed the networks with higher SNRs. A subset of SNRs used in training is presented in Figure \ref{transfer} (step one with simulated noise) and figure \ref{livingston} (step two, with real noise). We used Google's deep learning library, \texttt{tensorflow} \cite{tensor} for the entire analysis.

\section{RESULTS}
In our analysis, we define Peak Amplitude (PA) of the denoised signal as the detection threshold. Our detection strategy of an unknown signal is as follows. The data segments of $4096$s duration are fed into the network as eight second long data segments, and the denoising is performed on all the data segments. The segment with the presence of a real GW will ideally have the highest PA, while the PA of segments with no real GW event will be orders of magnitude lower than the PA of real GW. This is expected because the training of the Neural-Net was carried out in a  way to output zero if no GW was injected into the data stream. In each one hour long data, we decide the cut-off PA as the one with utmost one order of magnitude lower than the strongest signal (with the highest PA), and all other data segments with lower PAs are categorized as noise. In essence, the detection and denoising are performed in a single step, which is the separation of signal from the noise. In effect, the PA of the denoised signal will act as the SNR detection threshold of a matched-filter based detection.

Matched-filter based analysis defines the False Alarm Rate (FAR) of each detection as the number of false positive detections with an equal or higher ranking statistic, where ranking is assigned to each positive trigger that passes the SNR threshold of the detection pipeline and inter-site travel time requirement \cite{gwtc1}. While the FAR threshold used in $O3$ is $< 2/\text{day}$, most detections have an FAR that is of the order of years. This is orders of magnitude smaller than the false alarms that are typical to neural networks. In \cite{far}, it is noted that the FARs of matched-filter pipelines do not directly translate into the False Alarm Probability (FAP) characteristic of ML based GW search. In our analysis, we define the FAP based on the Peak Amplitude (PA) of a denoised GW signal. We define the FAP of a given detection as the ratio of number of denoised signals with a PA greater than or equal to the PA of a real GW signal. In the case of real events reported in this paper, FAP of an event is calculated using several hours of data before and after the event. This is calculated separately at both detectors, and are presented as `Single detector FAP' in table \ref{table}. The single detector FAPs can be further improved by analysing more data around the event and even beyond, hence it is not necessarily a fixed number, nor can it be elevated as a detection standard. Two-detector FAPs are defined as the false alarms from the same data segment at the same time at both the detectors. The two detector FAPs are found to be zero for all real GW events detected in our analysis. This means, no false alarms are produced under the condition that the detection needs to be registered at both the detectors at the same time. We used 10,000 data segments of eight second duration to calculate the FAPs presented in table \ref{table}. The PA of the real GW event is typically the highest among the analyzed data, and occasional false positives are counted towards the FAP of the event. 

The overlap between the template signal $h$ and the denoised signal $h^d$, with $N$ data points is calculated using the equation \cite{ml-gw}:

\begin{equation}
    O = {\sqrt\frac{\sum_{i=0}^{N} h_i h_i^d }{\sum_{i=0}^{N} h_i h_i}}
\end{equation}

\begin{figure*}
\centering
\includegraphics[width=9cm,height=6.5cm]{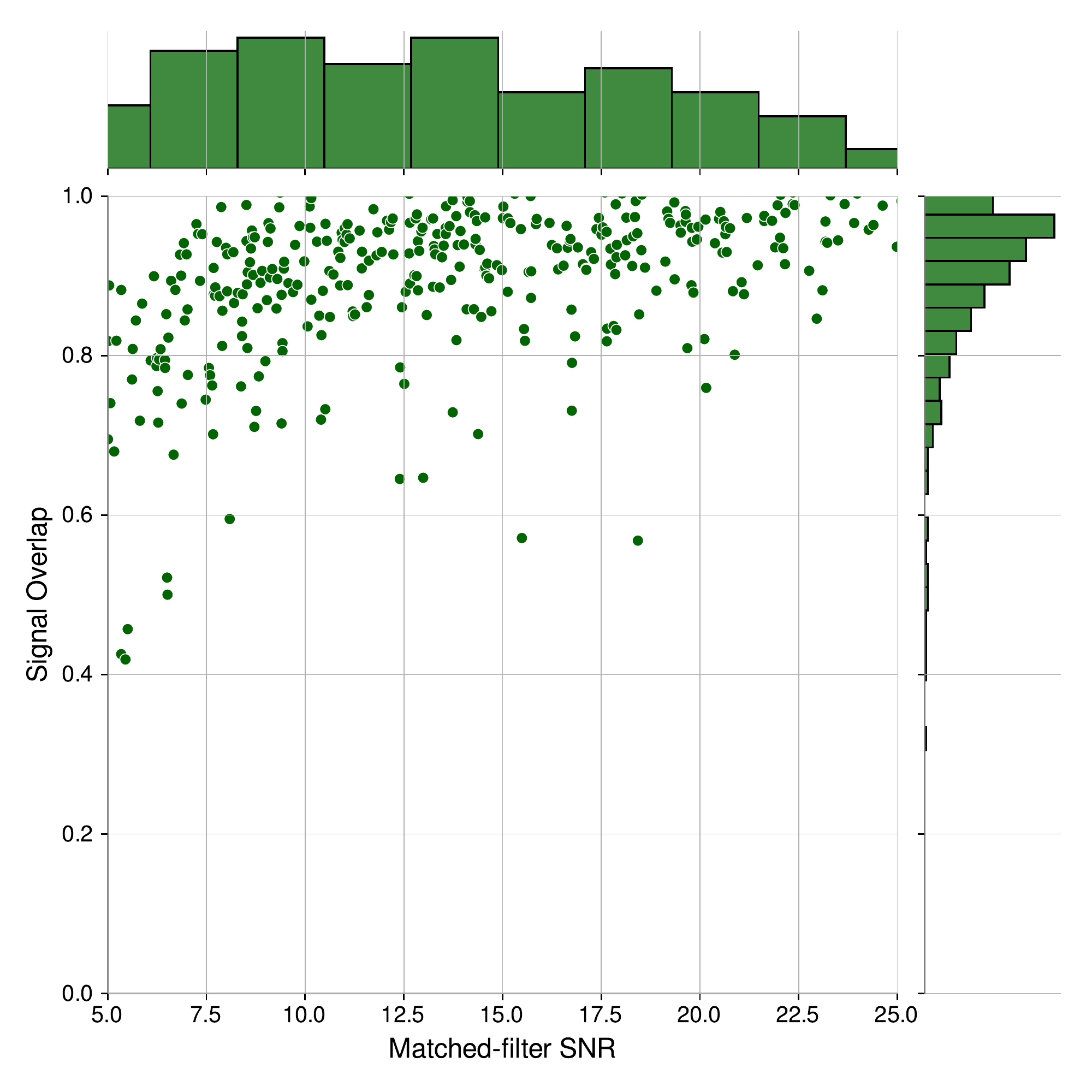}
\caption{Signal overlaps after denoising from the simulated noise alone, as function of the matched-filter SNR (before denoising). Top histogram shows the population sample of SNRs used for the analysis, side histogram shows the statistics of signal overlaps.}
\label{transfer}
\includegraphics[width=9cm,height=8cm]{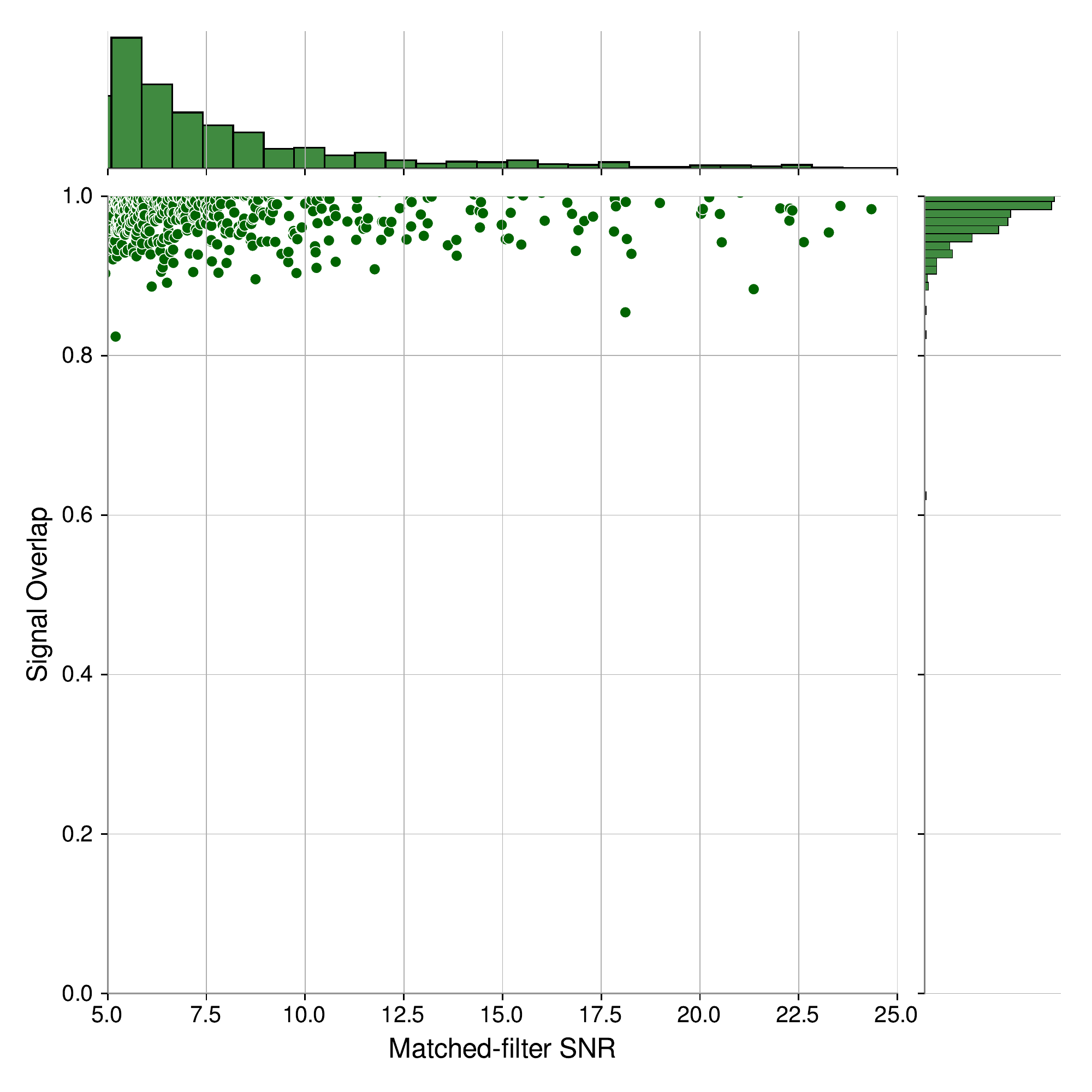}
\includegraphics[width=9cm,height=8cm]{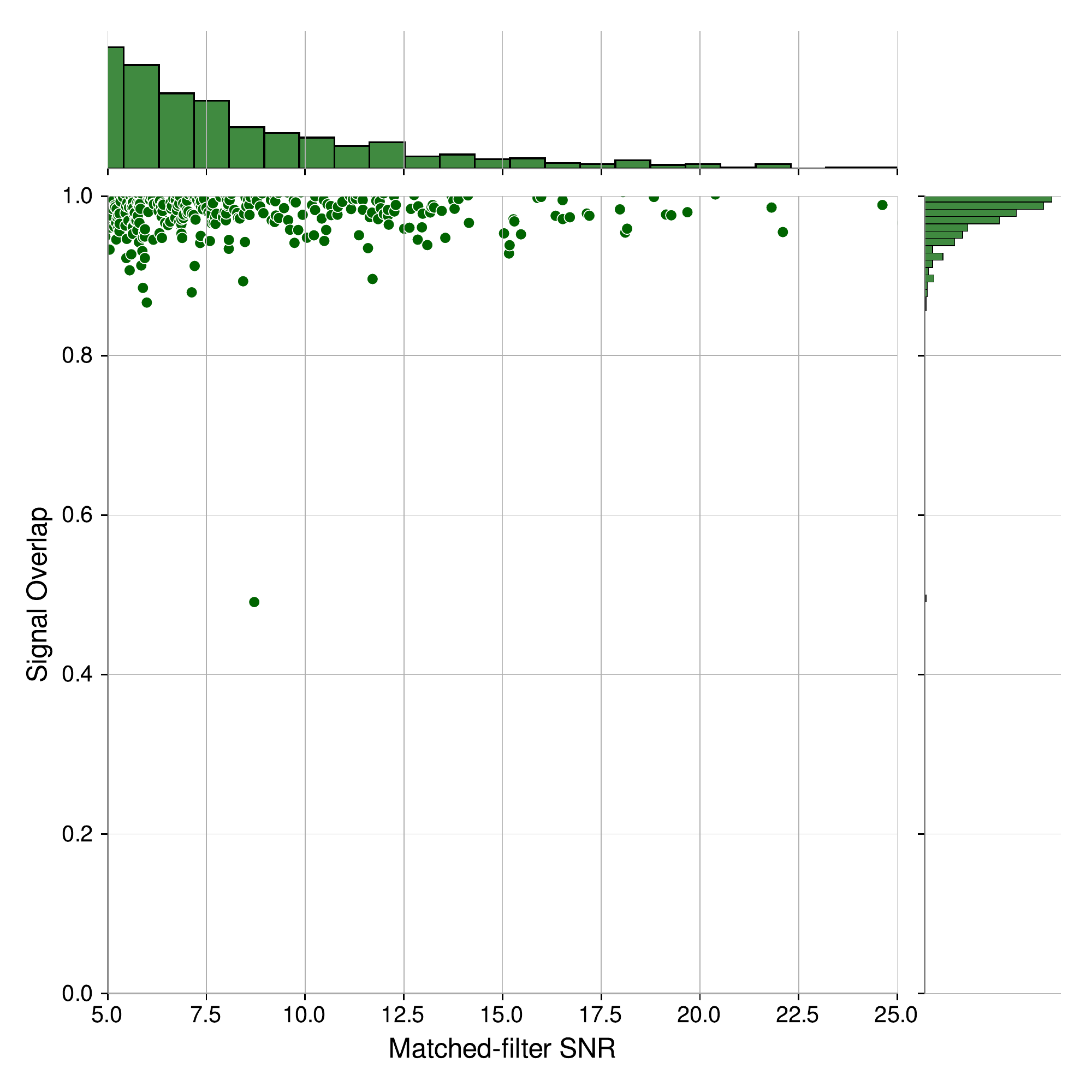}
\caption{Signal overlaps after denoising, at Hanford (left) and Livingston (right) as function of the matched-filter SNR (before denoising). Top histograms show the population sample of SNRs used for the analysis, side histogram shows the statistics of signal overlaps.}
\label{livingston}
\end{figure*}

Figure \ref{transfer} shows the overlaps of signals after denoising, as a function of the matched-filter SNRs before denoising, with simulated noise. Figure \ref{livingston} shows the same with real aLIGO noise from both detectors. The increase in signal overlaps between simulated noise and real noise can be seen, which is a result of transfer learning from the simulated noise analysis.

\subsection{\textit{GW150914}}
\begin{figure*}
\includegraphics[width=\textwidth, height=11.5cm]{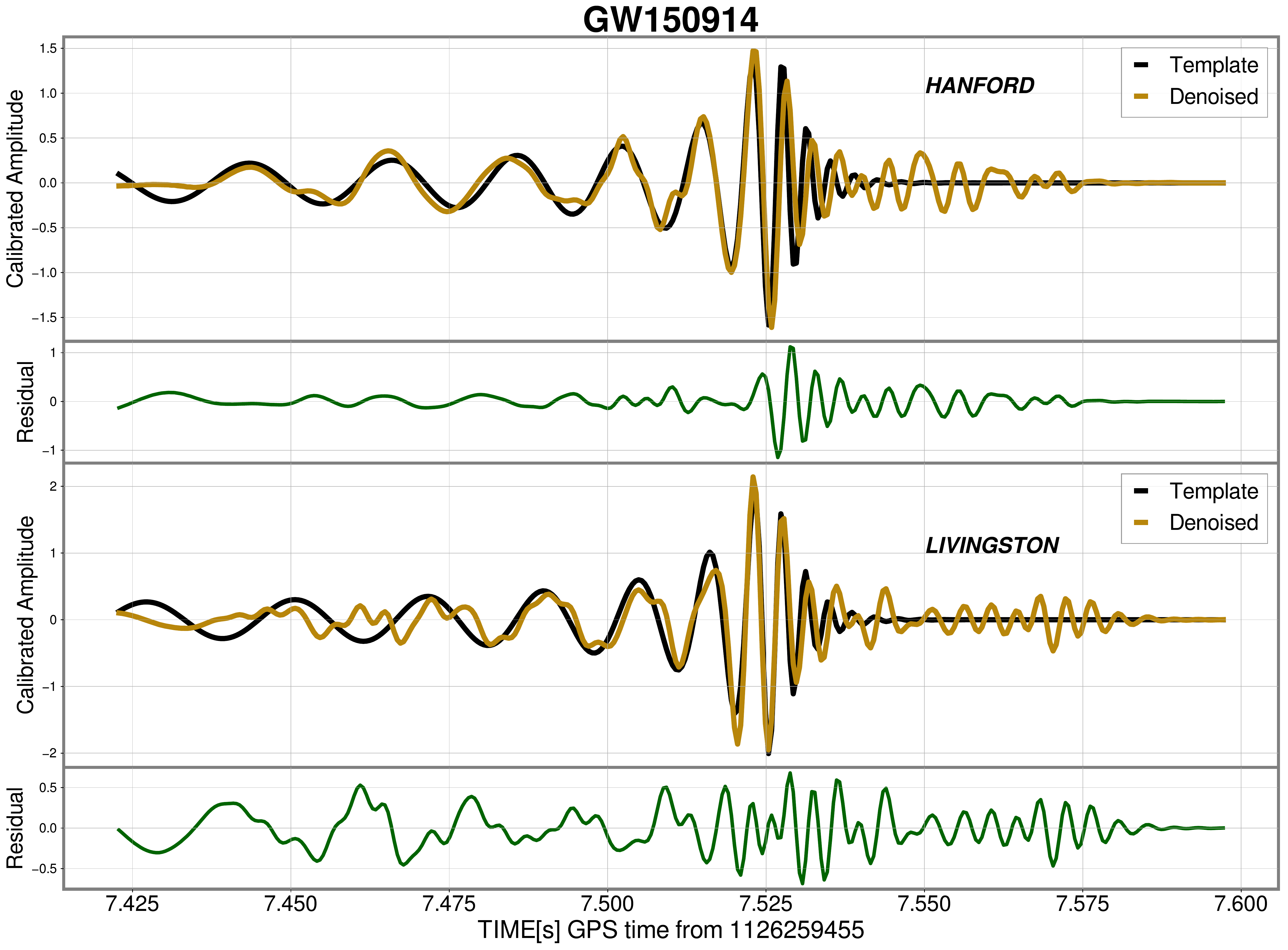}
\caption{\label{150914}GW150914 template and denoised waveform at Hanford (top panel) and Livingston (bottom panel). Amplitude scale is calibrated to match with the signal template}
\end{figure*}

In both detectors, GW150914 was identified as the signal with the highest PA at the exact time as the real detection. This merger of black holes of masses $35.6M\textsubscript{\(\odot\)}$ and $30.6M\textsubscript{\(\odot\)}$ was detected by all three search pipelines operational during $O1$ (PyCBC, GSTLaL and cWB) with a network SNR of $26$. We analyzed data, several hours before and after the merger event to calculate the FAP of the detection, the FAP at Hanford is estimated to be $0.01 \%$ and the FAP at Livingston is estimated to be $0.2 \%$. The denoised signals from both detectors are shown in Figure \ref{150914}. The overlap between the template of GW150914  which is simulated using the parameters estimated by Bayesian analysis by aLIGO \cite{properties} and the denoised signal is calculated using the signal overlap formula. Denoised GW150914 at Hanford matches very closely with the model template with $86 \%$ overlap. At Livingston, the model template matches $80 \%$ with the denoised signal. The better denoising performance at Hanford is most likely the result of the higher SNR of the detection at Hanford for this event, although this trend of higher signal overlap on account of higher SNR is not observed for all the events, hence cannot be generalized. For this event, we recovered a few cycles of inspiral phase followed by the merger and ringdown at both detectors. At Livingston, the denoised event is not exactly in phase with the template at early part of the inspiral, but later on it comes to be in phase. At both detectors, the ringdown phase appears to be longer and also appears to be deviating from the signal template at the final stage of ringdown. This trend of `ringing' after ringdown is observed for many of the recovered signals, although this requires further analysis to infer if it is indeed a real physical phenomena or an unavoidable feature of the denoising scheme. The residuals are shown in separate panels for both detectors, which arise primarily due to the amplitude mis-match and phase differences between the template and the denoised signal, as opposed to the presence of unsubtracted noise. The frequency content of the residual is similar to the frequency content of the event itself, at all three phases event, which also is a consequence of the direct signal extraction, characteristic of the denoising scheme.  

\subsection{\textit{GW170104}}
\begin{figure*}
\includegraphics[width=\textwidth,height=10cm]{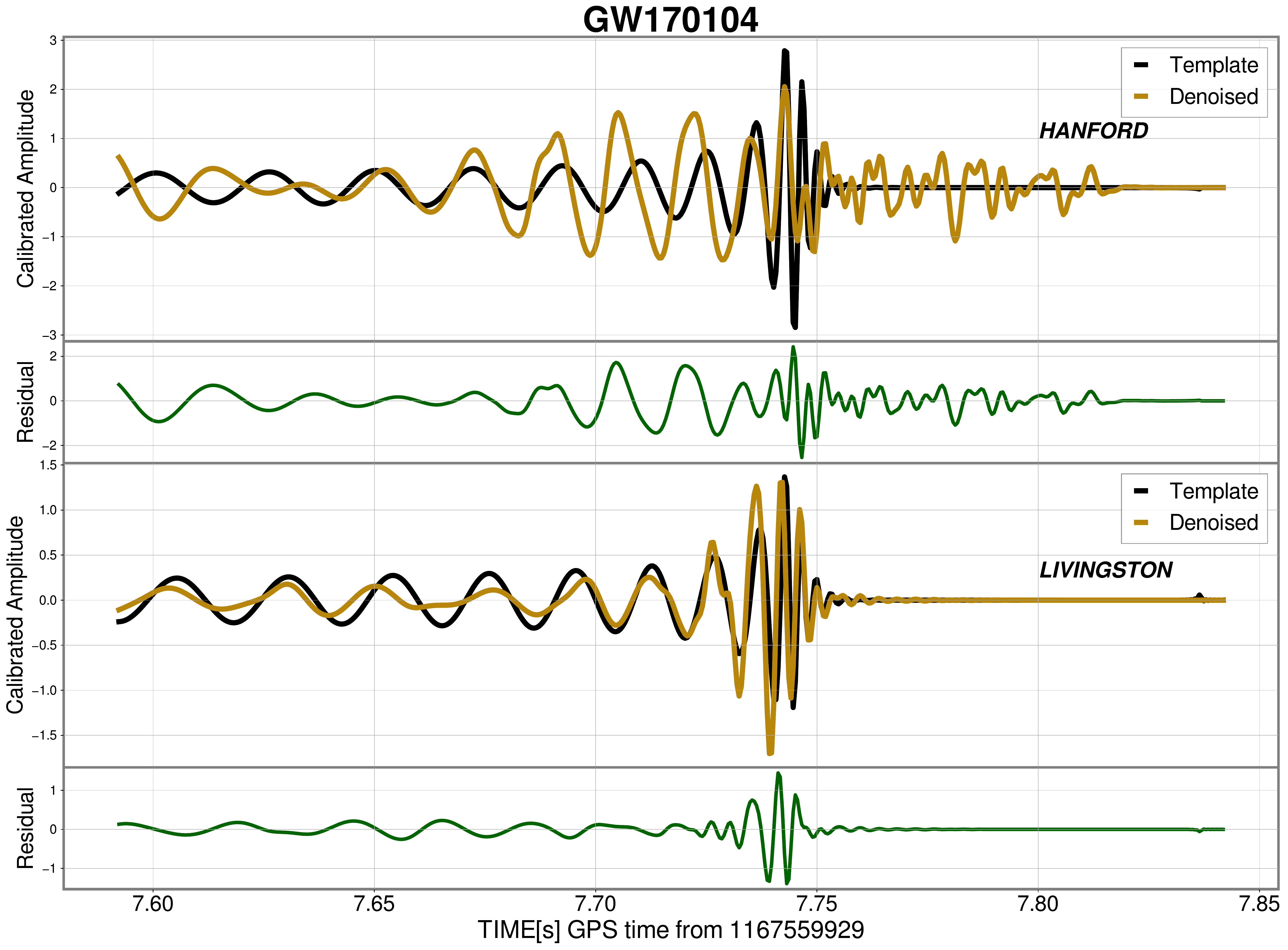}
\caption{\label{170104}GW170104 template and denoised waveform at Hanford (top panel) and Livingston (bottom panel). Amplitude scale is calibrated to match with the signal template.}
\end{figure*}

GW170104 is a BBH merger of masses $30.8M\textsubscript{\(\odot\)}$ and $20 M\textsubscript{\(\odot\)}$, detected by all three search pipelines with a network SNR of $13.8$. At Hanford, the denoised signal is out of phase with the template at the early phase of inspiral and the ringdown phase deviates from the template and is longer, with an overall signal overlap of $77\%$. At Livingston, the denoised signal matches perfectly with the template at all three phases of the merger, with an overlap of $85\%$. The single detector FAPs are estimated to be $0.4\%$ and $0.15\%$ respectively and the two detector FAP is zero. See Figure \ref{170104}.

\subsection{\textit{GW170809}}
\begin{figure*}
\includegraphics[width=\textwidth,height=10cm]{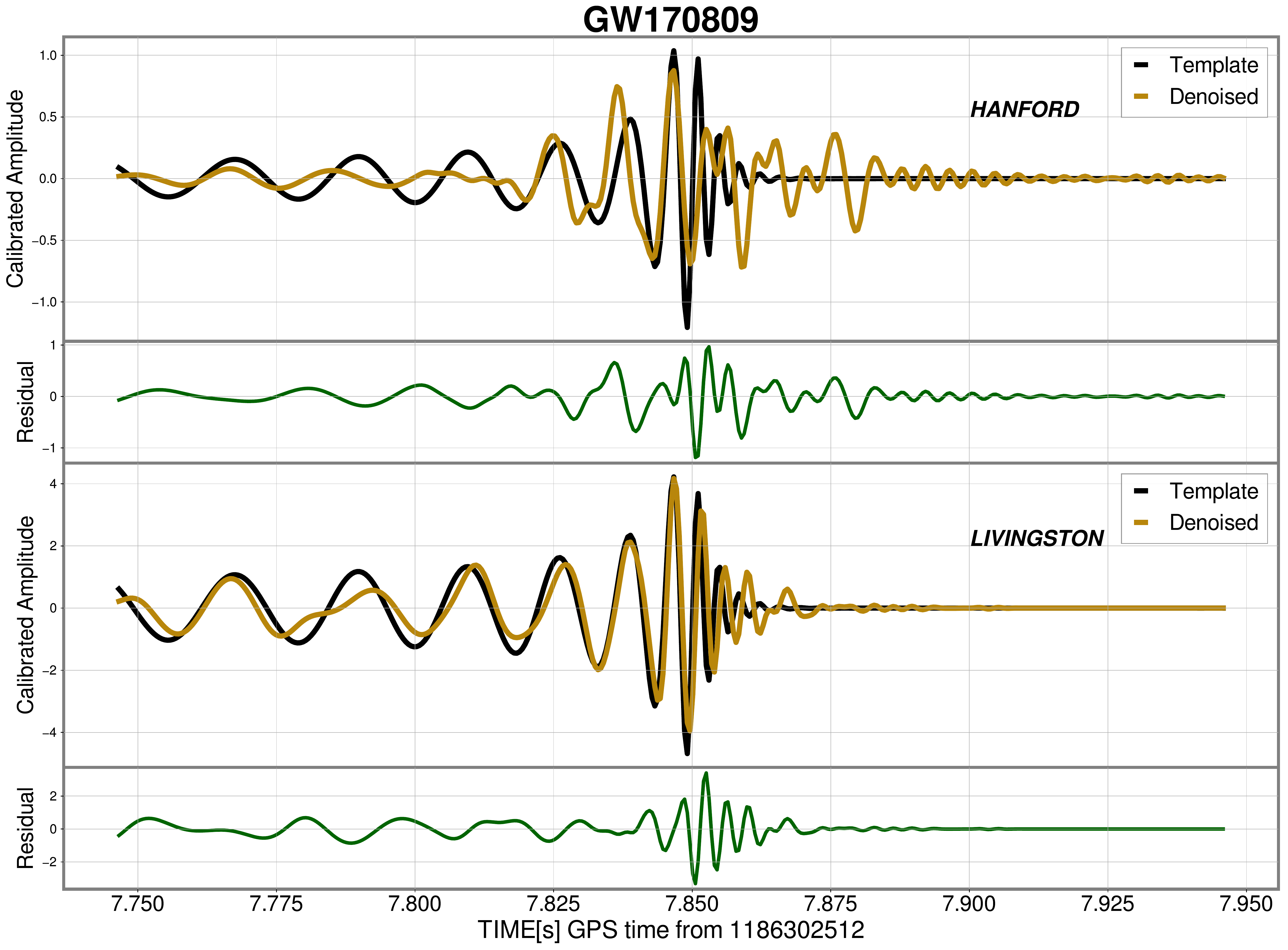}
\caption{GW170809 template and denoised waveform at Hanford (top panel) and Livingston (bottom panel). Amplitude scale is calibrated to match with the signal template.}
\label{170809}
\end{figure*}

GW170809 is a BBH merger of masses $35M\textsubscript{\(\odot\)}$ and $23.8 M\textsubscript{\(\odot\)}$, detected by PyCBC and GSTLal pipelines with a network SNR of $12.8$. At Hanford, the signal overlaps fairly well with the template at late stages of inspiral and merger, with an overall overlap of $68\%$. The ringdown phase deviates from the template towards the end and displays the `ringing' after ringdown, which sustains for a few milli seconds. At Livingston, the signal overlaps very well with the template at all three phases of the merger, with an overlap of $85\%$. Single detector FAP at both detectors are estimated to be about $0.05 \%$ and two detector FAP is zero. See Figure \ref{170809}.

\subsection{\textit{GW170814}}

\begin{figure*}
\includegraphics[width=\textwidth,height=10cm]{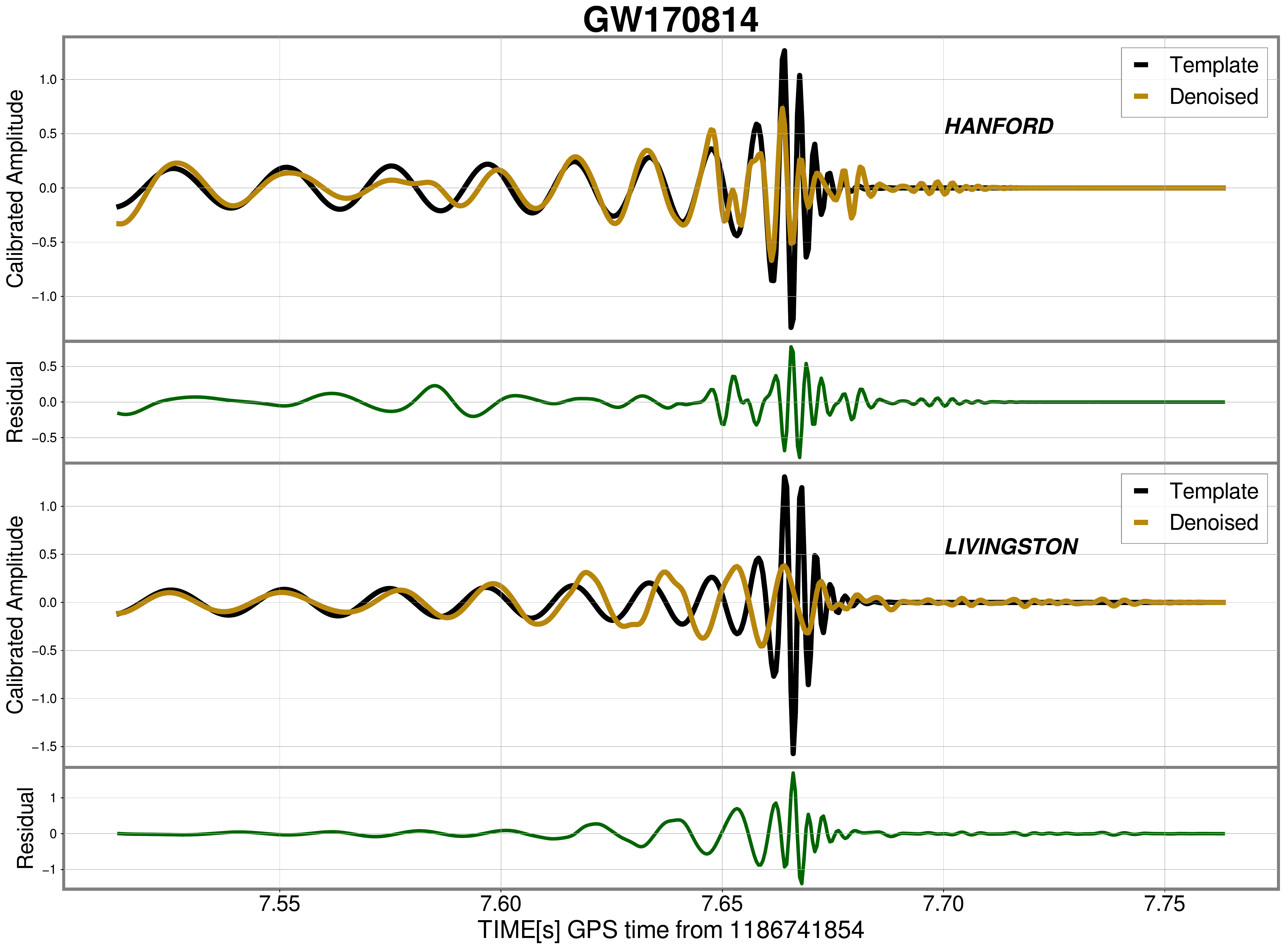}
\caption{GW170814 template and denoised waveform at Hanford (top panel) and Livingston (bottom panel). Amplitude scale is calibrated to match with the signal template.}
\label{170814}
\end{figure*}

GW170814 is a BBH merger of masses  $30.6M\textsubscript{\(\odot\)}$ and $25.2 M\textsubscript{\(\odot\)}$, detected by all three search pipelines with a network SNR of $17.7$. At Hanford, the denoised signal overlaps well with the template at all three phases with an overall signal overlap of $90\%$. There appears to be a 'ringing' immediately after the ringdown phase. At Livingston, the signal overlaps well only at early inspiral phase, and the overall signal overlap is only $41\%$.  The denoised merger does not recover the highest frequencies at the merging phase, but the frequency is close at the ringdown phase, which appears to persist longer than the template ringdown. The single detector FAPs are estimated to be $0.1\%$ and $0.25\%$ respectively and the two detector FAP of the detection is zero. See Figure \ref{170814}.

\subsection{\textit{GW170729}}
\begin{figure*}
\includegraphics[width=\textwidth,height=10cm]{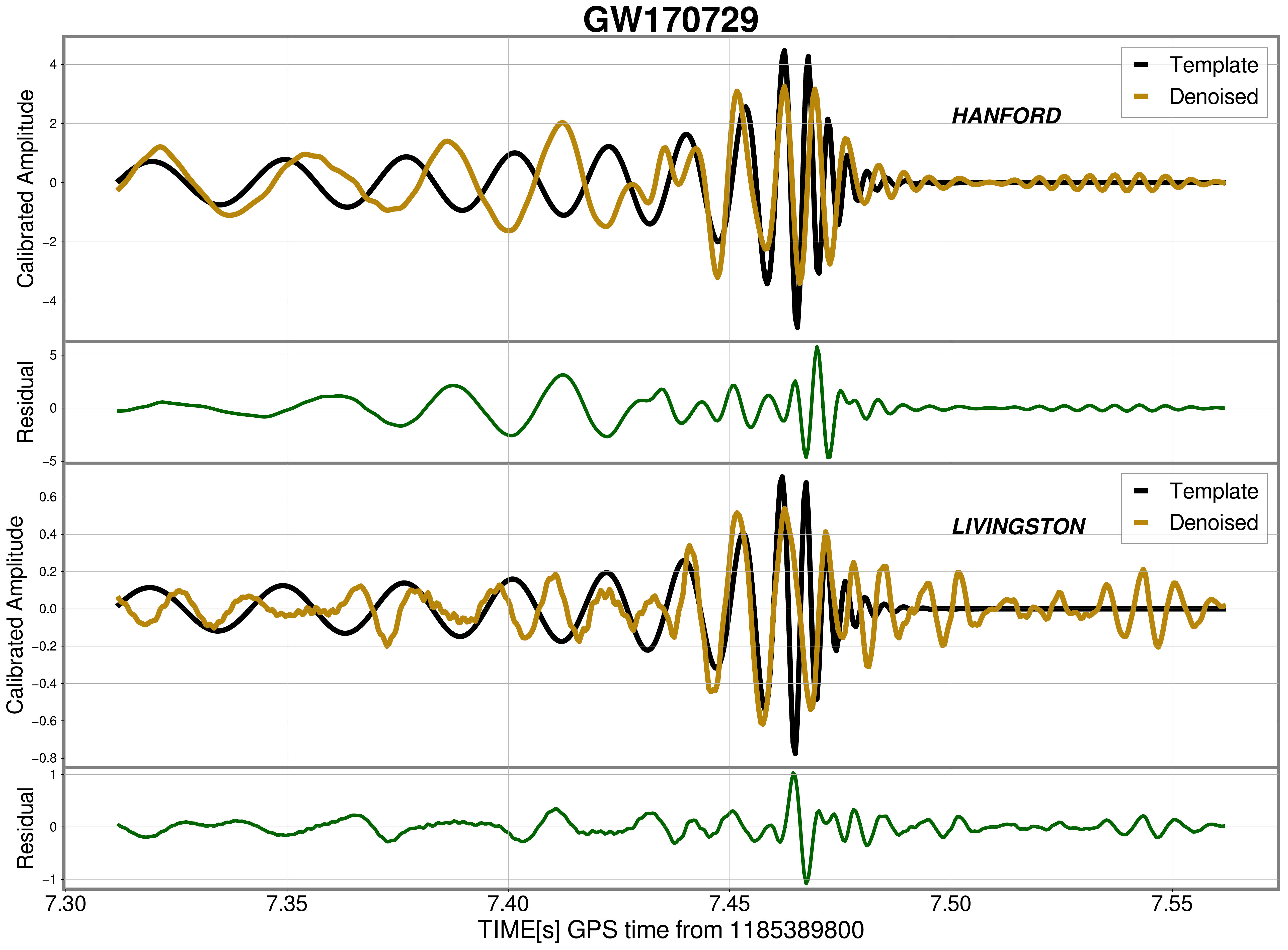}
\caption{GW1700729 template and denoised waveform at Hanford (top panel) and Livingston (bottom panel). Amplitude are calibrated to match with the signal template.}
\label{170729}
\end{figure*}

GW170729 is BBH merger of masses $50.2M\textsubscript{\(\odot\)}$ and $34 M\textsubscript{\(\odot\)}$, detected by all three search pipelines with a network SNR of $10.8$. This is the heaviest system detected in $O2$. The denoised signal has the same frequency at all three phases at the Hanford detector, although out of phase at some portion of the inspiral and the overall signal overlap is calculated as $60\%$. The extra 'ringing' after the ringdown is clearly visible. At Livingston, the inspiral part largely has the same frequency as the template, although there appears to be higher frequencies riding on top of the lower frequencies, which is observed only for this event. The waveform is in phase with the template at late inspiral and merger. The overall signal overlap is $67\%$. The ringing after ringdown is visible, with a higher amplitude than the Hanford detector. Single detector FAPs are $0.6\%$ and  $4\%$ respectively and the two detector FAP is zero. See Figure \ref{170729}.

\subsection{\textit{GW170608}}
\begin{figure*}
\includegraphics[width=\textwidth,height=10cm]{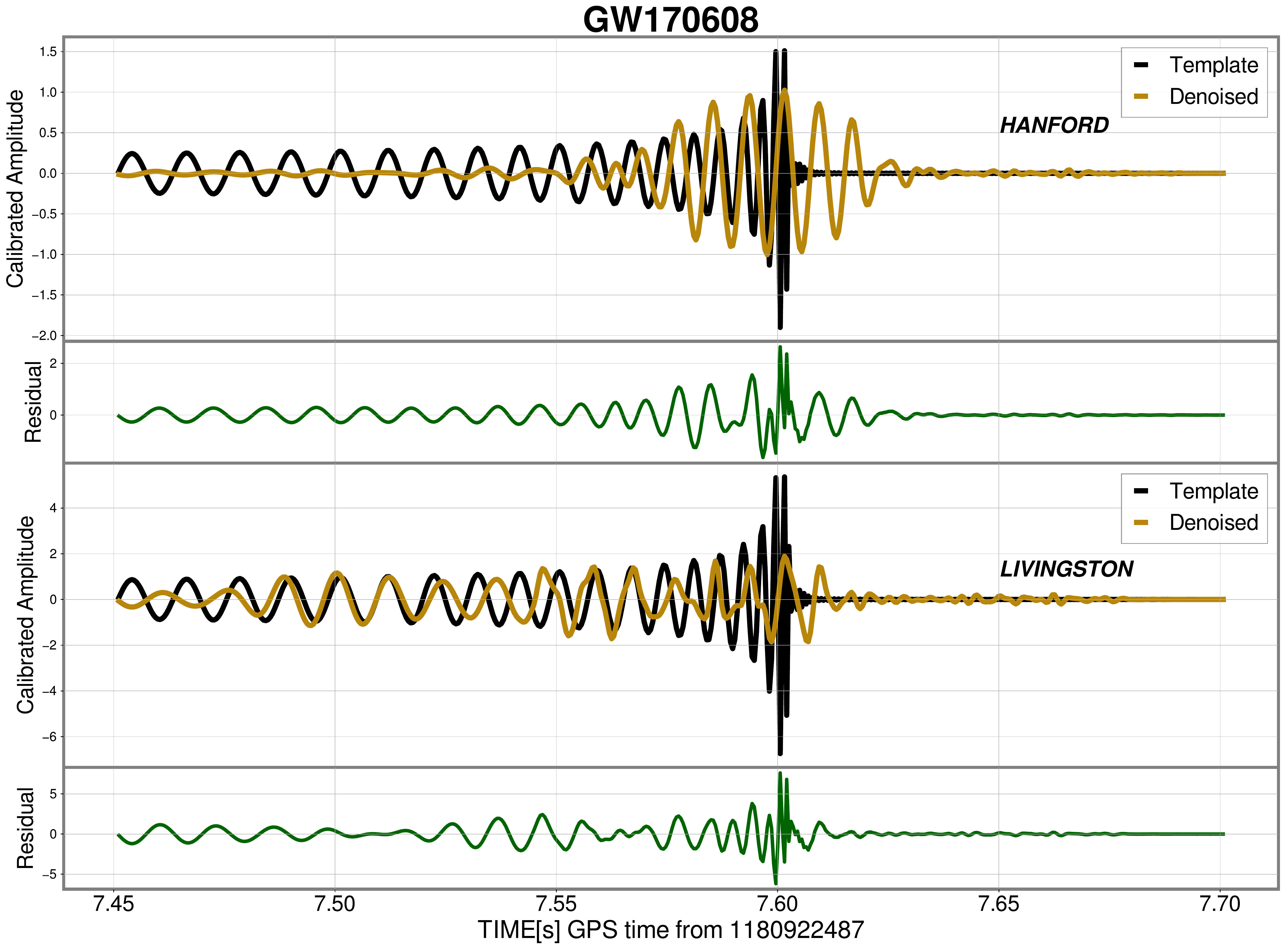}
\caption{GW170608 template and denoised waveform at Hanford (top panel) and Livingston (bottom panel). Amplitude scale is calibrated to match with the signal template.}
\label{170608}
\end{figure*}

GW170608 is a BBH merger event of masses $11M\textsubscript{\(\odot\)}$ and $7.6 M\textsubscript{\(\odot\)}$, detected by all three search pipelines with a network SNR of $15.4$. This is the lightest event of O2 with the highest frequencies at all three phases. At both detectors, the denoised signals deviate from the template as the Neural-Net fails to recover the highest frequencies at the merger and ringdown, characteristic of the event. At both detectors, the merger appears to be longer than what the template suggests and the ringdowns arrive late, and persists longer than the template. The signal overlaps are only $31\%$ and $48\%$ at Hanford and Livingston, and the single detector FAPs of $4\%$ and $0.1\%$ respectively. The two detector FAP of the detection is zero. See Figure \ref{170608}.
 
\subsection{\textit{GW170823}}
GW170823 is a BBH merger event of masses $35.8M\textsubscript{\(\odot\)}$ and $29 M\textsubscript{\(\odot\)}$, detected by all three search pipelines with a network SNR of $12.2$. At Hanford, the denoised signal overlaps well with the template at the inspiral phase but does not recover the highest frequencies at the merger. The ringdown deviates from the template and lasts longer and the overall signal overlap is about $60\%$. At Livingston, the signal overlaps with template well at inspiral phase and reasonably well at merger and ringdowns. The overall signal overlap is calculated to be of only $30\%$. The single detector FAPs are $2\%$ and $1\%$ respectively and the two detector FAP of the detection is zero. See Figure \ref{170823}.

\subsection{\textit{GW170818}}
\begin{figure*}
\includegraphics[width=\textwidth,height=7cm]{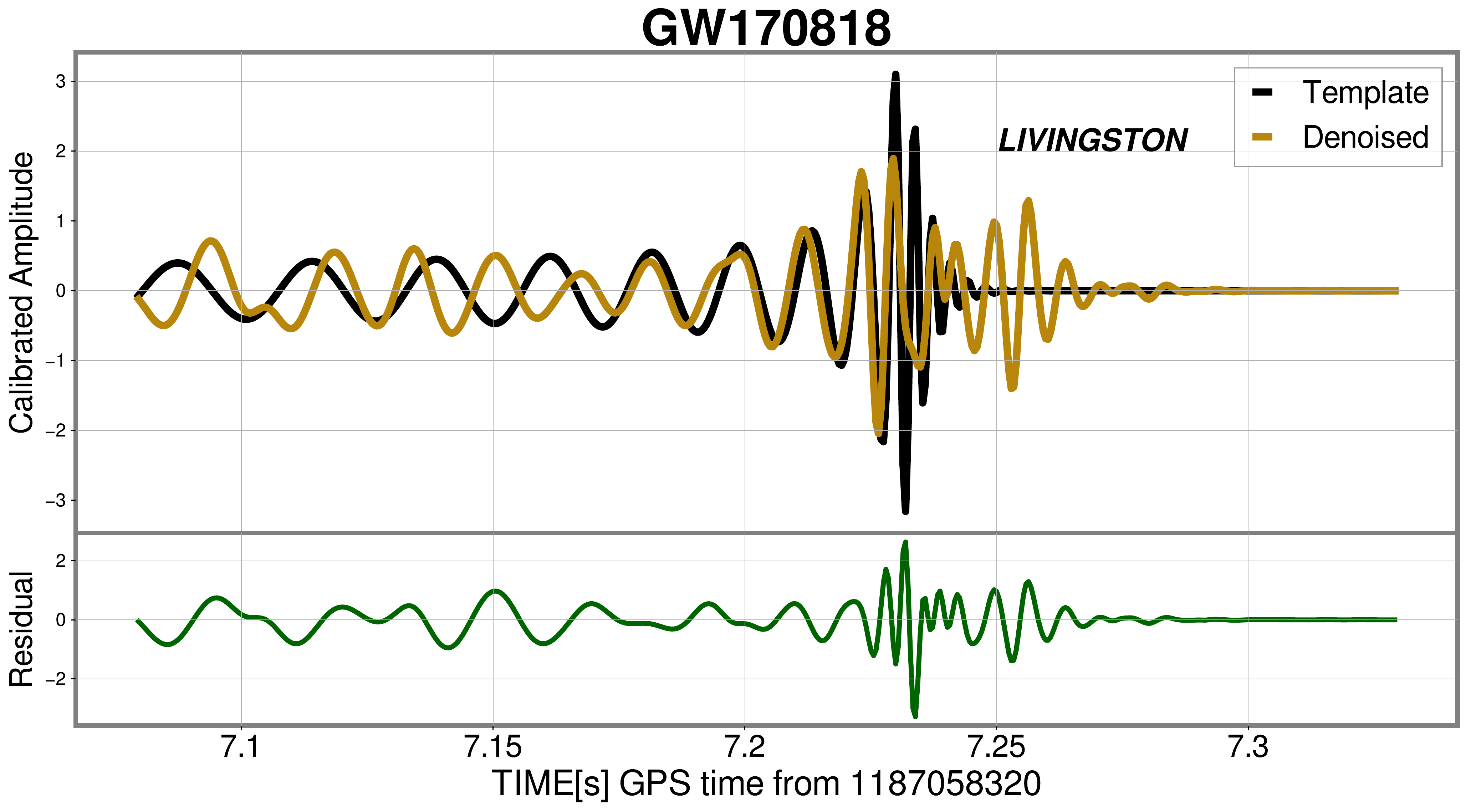}
\caption{GW170818 template and denoised waveform at Hanford(top panel) and Livingston (bottom panel). Amplitude scale is calibrated to match with the signal template.}
\label{170818}
\end{figure*}

GW170818 is a BBH merger of masses $35.4M\textsubscript{\(\odot\)}$ and $26.7 M\textsubscript{\(\odot\)}$, detected by only one of the search pipeline (GstLAL) with a network SNR of $12$. The detection of this event was extremely weak at the Hanford detector with an SNR of only $4.6$, which is only slightly above the detection threshold of both GstLAL and PyCBC search pipelines. Our Neural-Net detected this event only at the Livingston detector, and could not reliably provide a peak amplitude at the Hanford site, hence not included in this article. At the Livingston detector, the frequency largely matched with the template at the inspiral phase, although not in phase with the template at early inspiral. The merger and ringdown phases are in phase with the template and there is an additional `ringing' immediately after the ringdown, but not as long as many of the other denoised signals. The signal overlap and FAP are $70\%$ and $5\%$ respectively. 
See Figure \ref{170818}.

\begin{table*} 
\caption{ \label{table} Summary of the real GW events analysed from both aLIGO detectors. Table shows the optimal matched-filter SNRs, single detector FAPs and signal overlaps.}
\begin{ruledtabular}
\begin{tabular}{cccccccc}
&\multicolumn{1}{c}{}&\multicolumn{2}{c}{Matched-filter SNR}&\multicolumn{2}{c}{Single detector FAP[$\%$]}&\multicolumn{2}{c}{Signal overlaps[$\%$]}\\
 Event&GPS&Hanford&Livingston&Hanford&Livingston&Hanford&Livingston
\\ \hline
 GW150914&$1126259462.4$&$20.6$ &$14.2$&$0.01 $&$0.2  $&$86.1$&$80$\\ 
 \\
 GW170814&$1186741861.5$&$9.3$&$14.3$&$0.1 $&$0.25 $ &$90$&$41$\\
\\
 GW170104&$1167559936.6$&$9.5$&$9.9$&$0.4$&$0.15  $&$77$&$85$\\
 \\
 GW170823&$	1187529256.5$&$6.8$&$9.2$&$2 $&$1$&$60$&$30.5$\\
 \\
 GW170809& $1186302519.8$&$5.9$&$10.7$&$0.05$ &$0.05$&$68.8$&$85.6$\\
 \\
 GW170608&$1180922494.5$ &$12.1$&$9.2$&$ 4  $&$0.1  $&$31$&$48$\\
 \\
 GW170729&$	1185389807.3$ &$5.9$&$8.3$&$0.6$&$4 $&$60$&$67.2$\\
 \\
 GW170818&$1187058327.1$ &$4.6$&$9.7$& $-$&$5 $&$-$&$70$\\
\end{tabular}
\end{ruledtabular}

\end{table*}

\section{Discussion}

The results presented in this article are highly encouraging for this innovative CNN technique. This optimization scheme is demonstrated to have the ability to detect and separate GW signals from highly non-stationary and non-Gaussian noise, hence warrants further exploration.

\begin{figure*}
\includegraphics[width=\textwidth,height=10cm]{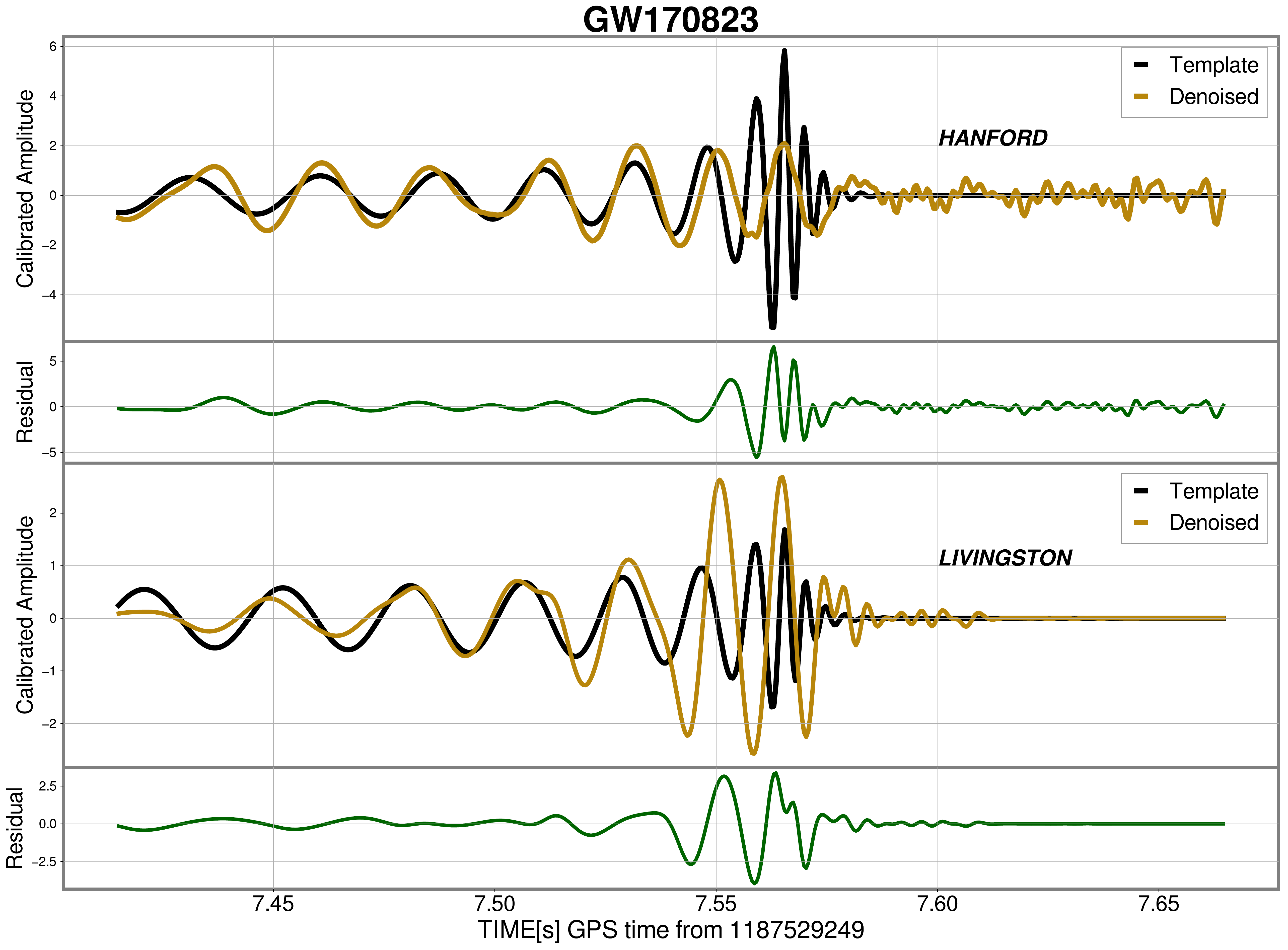}
\caption{GW170823 template and denoised waveform at Hanford (top panel) and Livingston (bottom panel). Amplitude scale is calibrated to match with the signal template.}
\label{170823}
\end{figure*}

The signal non-overlap of some of the detected events with the model template is possibly due to the absence of spin effects that is missing in this analysis, especially for the events where the denoised signal is missing a complete cycle, in the inspiral. Spin projections along the direction of orbital angular momentum affect the inspiral rate of the binary. The spin components aligned with the orbital angular momentum increase the number of orbits while the spin components anti-aligned with the orbital angular momentum decrease the number of orbits from any given separation to the merger with respect to non-spinning case \cite{spin1, spin2}. In some of the denoised waveforms, we see certain cycles partially missing, or attenuated in amplitude (at $7.56$s and $7.57$s of GW170823 - Figure \ref{170823}) rapidly, without reconstructing the complete cycle. This is more likely an effect related to the CNN itself, as opposed to a physical effect, and can in principle be tackled by further optimizing the hyper-parameters of the Neural-Net or by incorporating some of the latest features developed in various deep learning libraries. The two-detector FAP is zero for all the detected events, which is promising from the detection and early alert point of view.  

There are a multitude of ways to improve this detection strategy. Firstly, this analysis is two dimensional. The only source property that is used in characterizing the signals used for training the Neural-Net is the black hole masses. Although ML methods are equipped with the capacity to interpolate between waveform templates, we expect the performance of the Neural-Net to reliably get better by training with a few more source properties such as spins of individual black holes, inclination and  eccentricities. In addition, the training was exclusively based on one family of NR waveforms (SEOBNRv4), which can be expanded with more NR waveform families with higher modes and precession effects included, such as IMRPhenomXPHM \cite{imr} and SEOBNRv4PHM\cite{seo}. This extended analysis is currently underway and will be communicated in the future. Enhancing the parameter space with more properties of the binary does not compromise on the fast performance of this method to detect and denoise GW, since all the intensive computations are performed during the training stage, which is a one time process. This is different from the matched-filter analysis, where the computations need to be repeated over and over again for the entire observing run. Although the Neural-Net is expected to have high fidelity to search for real events after being trained on noise from a previous observing run, the best results can be expected by retraining the network with the current observing run, as the sensitivity of the detector is being improved constantly, and hence the characteristics of the noise is changing over time. Another area of interest is, expanding this formalism to include BNS and NSBH mergers.

One of the plausible explanations for the 'ringing' phenomena could be the proposed 'echos' arising from the introduction of structures near the event horizon of black holes. These echos are late, repeated ringings of the ringdown phase of a black hole merger as a result of waves trapped between the near-horizon structure and the angular momentum barrier, as first discussed in \cite{echo1, echo2}. The observational evidence of echos in a real GW event is further discussed and debated in \cite{echo3, echo4, echo5}. It requires a careful analysis of the characteristics of these patterns and comparison to the theoretically proposed features of the echos to confirm if these are indeed observational signatures of the echos. 

\section{CONCLUSION}

In this article we presented a deep learning framework and its potential to  detect and denoise GW signals from black hole binaries. The network learns a sparse representation of the data and separates signal from noise by generating two adaptively thresholding masks. Effectively, we are subtracting noises from all frequency bands including where the frequency content of the noise overlaps with that of the signal. From the unprocessed raw strain data of the detector, the Neural-Net successfully detected all the black hole binary mergers from the second observing run of aLIGO. This ML based detection strategy is a strong candidate to be incorporated into the search, analysis and parameter estimation of merger events in the upcoming observing runs of ground based detectors. Understanding the nature of various complex noise sources and studying them statistically can be extremely challenging in the context of GW detectors, hence this task can virtually be outsources to deep neural networks which are equipped to learn the patterns that are characteristics of the noise themselves.  It is our view that a Neural-Net based detection and denoising will work best in conjunction with the matched-filtering method, for both detection and parameter estimation.

With the prospect of detecting hundreds of GW events in the nearby future with current and emerging ground based detectors, analyzing months of detector data output is a very challenging task. A deep learning search pipeline can be immensely helpful in faster detections, efficient data processing and increase/reduce the confidence of a given detection. An analysis of marginal detections \cite{gwtc2.1} using ML methods is also crucial, which can help confirm or reject the marginal detection, in addition to looking for potentially `missed' real GW signals. 

The transfer learning technique that we adopted to improve the performance can be used to retrain the network with data from the most recent observing run. Retraining or transfer learning the network with new data is fairly easy given the computational resources available today that are exclusively powerful for ML based analysis. Retraining this network takes only a few minutes in a GPU based HPC system. Also, the network can always be updated with new template signals developed that are potential candidates including other astrophysical events beyond CBC.

The online, low-latency search for events which are common place during the observing runs of aLIGO can be made more efficient by incorporating a Neural-Net based early alert for a multi-messenger counterpart. The emerging hardware and software infrastructure of AI and GPU based parallel computing are on an accelerated trajectory today, which are very promising developments for the GW astronomy in the coming years.

\begin{acknowledgments}
The authors would like to thank the computational resources granted for this research by the Texas Advanced Computing Center (TACC - project AST22014) and Optane cluster resource provided by the Seismology research group (David Lumley, Hejun Zhu) of University of Texas at Dallas. C. M is supported through the research assistantship program by David Lumley.

\end{acknowledgments}


\nocite{*}

\clearpage

\bibliography{PRD}

\end{document}